\documentclass[10pt]{iopart}
\usepackage{iopams}
\usepackage{bm}  
\usepackage{graphicx}

\begin{document}

\title[Madden-Julian Oscillation described as a Burgers Kink]{Madden-Julian Oscillation described as a Nonlinear Burgers Kink in the Meridional Vorticity Equation}

\author{Richard Blender}
\address{Meteorological Institute, Universität Hamburg, Hamburg, Germany}
\ead{richard.blender@uni-hamburg.de}

\begin{abstract}
A dynamic equation for a large scale convective event in the tropical atmosphere similar to the Madden--Julian Oscillation (MJO) is suggested based on the meridional vorticity equation with buoyancy parametrized by Convective Available Potential Energy (CAPE). The propagation is determined by the nonlinear Burgers equation with a stationary solution describing a kink moving towards the moisture source. In this conceptual model, the propagation speed depends on the asymmetry in the zonal surface winds which is observed in the boundary layer in the vicinity of an MJO event and attributed to Rossby and Kelvin waves.  Furthermore, the model predicts convection at the equator in the east of the MJO which is not correlated with Kelvin and Rossby waves. 
\end{abstract}

\vspace{0.5cm}

\vspace{0.5cm}
\begin{center}

%preprint
%\today

\end{center}

\maketitle

%=======================================================  
\section{Introduction}\label{sec_Intro}
%=======================================================  

% Madden Julian

The Madden--Julian Oscillation (MJO) \cite{madden1971detection} appears as a large scale and intense precipitation field in the Indian ocean with a huge global impact on weather and climate \cite{zhang2005madden}.  The MJO can be projected on well-known linear tropical waves \cite{matsuno1966quasi,vzagar2015systematic} 
but has a much lower progression speed,  $c \approx 5 \, \mbox{ms}^{-1}$. Furthermore, whereas the major part of tropical precipitation is related to these waves, a considerable contribution close to the equator is not explained  \cite{zhang2012potential}. 

Although the MJO is the dominant part of the intraseasonal (30--90 days) variability in the tropics, the understanding  and modelling by GCMs  is considered incomplete \cite{ahn2017mjo}. While linear dynamics with heating   is well understood \cite{gill1980some}, there is ongoing search for models representing main mechanisms and properties, see the review  \cite{zhang2020four}. 
% 
% note that all 37 GCMs assessed therein are based on the TA. 
% 
Whereas \cite{zhang2020four} considers mainly parametrized physical processes, the analysis of \cite{zhang2012potential} hints at a dynamical mechanism.  Possible reasons for the difficulties in the perception and the understanding of the MJO are discussed in 
\cite{MJOasaGestalt} from a philosophical perspective.

% Traditional 

The majority of weather forecast and climate models assumes hydrostatic conditions and neglects the horizontal meridional vorticity and the northward component of the Coriolis vector on the sphere (coined {\it Traditional Approximation} (TA) by \cite{eckart1961hydrodynamics}). Simplified models also based on the TA have been proposed which explain the MJO as a localised structure in nonlinear equations: \cite{khouider2012climate} derived a Korteweg-de\,Vries (KdV) equation with soliton solutions for resonant baroclinic and barotropic Rossby waves, \cite{yano2017tropical} suggested equatorial modon solutions and \cite{zhao2021equatorial} a modon in the moist-convective shallow water equation. A realistic but yet simplified MJO model is the stochastic skeleton model  \cite{Thual-Majda_2014} of a neutrally stable atmosphere  coupled to an ocean surface which describes interactions with the Walker circulation and ENSO on various time scales \cite{yang2021enso}. 

%
% Nontraditional 
%

Nontraditional models might solve long standing problems like, for example, the life cycle of convective systems with observed periods of two days in outgoing long wave radiation  \cite{chen1997diurnal,igel2020nontraditional}. 
Prospects for avoiding the TA in nontraditional models have been already pointed out by \cite{kasahara2003nonhydrostatic}, and \cite{hayashi2012importance} have underlined the relevance of the full Coriolis force for large scale tropical motions. 
Atmospheric dynamics with the full Coriolis force and the impact of approximations on conservation laws is described in-depth by \cite{white1995dynamically} and a review of the effects of the nontraditional approximation is given by \cite{gerkema2008geophysical}.
The traditional approximation was, at least partially, avoided  in the quasi-hydrostatic equation in the UKMO weather forecast model early in 1992 \cite{cullen1993unified}. 
At present there are steps towards including the  nontraditional Coriolis component in the upper atmosphere in the atmospheric model ICON  \cite{borchert2019upper} which is used by the German Weather Service and the Max-Planck Institute for Meteorology in Hamburg \cite{zangl2015icon}. 

At the equator a particular type of waves exists in the nontraditional equation for the meridional (northward) vorticity component in the Boussinesq approximation in the vertical $x$-$z$-plane  \cite{ong2020compressional}. These waves are coined compressional Rossby waves and propagate eastward with a slow phase speed $c=2\Omega/H$ with angular velocity $\Omega$ and scale height $H$ (for $H=9.1 \, \mbox{km}$,  $c = 0.24 \, \mbox{ms}^{-1}$).

The aim of this publication is to present a parsimonious model for the progression of a large scale convective event at the equator in the Boussinesq approximation with a nonlinear parametrization of buoyancy. The model is guided by the dynamics of compressional Rossby waves and therefore nontraditional, although the horizontal Coriolis force term is discarded after a scale analysis. We derive a dynamic equation for the first baroclinic mode of the meridional vorticity component and assume neutral stratification. Linear tropical waves like Kelvin and Rossby waves are not included.  

The scales in our model follow the observations in \cite{hsu2012role}, who pointed to the major role of boundary layer moisture for the MJO and found that the westward advection of moisture in the boundary layer is responsible for the eastward propagation of the Madden--Julian oscillation. We parametrize convection using  CAPE (Convective Available Potential Energy) to relate buoyancy to the large scale vorticity \cite{zhang2005effects,narendra2010seasonal}. Heuristically the nonlinear parametrization can be interpreted as the product of lower tropospheric moisture transport towards the MJO  times vertical velocity. 

We obtain a Burgers equation where nonlinear steepening is balanced by horizontal diffusion which is supposed to originate in fast gravity waves observed in the vicinity of deep convection \cite{Jewtoukoff_Gravitywaves_2013}.  The equation has a stationary nonlinear solution with a step-like kink in the vorticity which shows unidirectional eastward propagation towards the moist boundary layer with a speed related to the zonal asymmetry. In \cite{Wang-MJOPropagation} this asymmetry is found in observations and GCMs with a realistic MJO propagation speed and attributed to stronger Kelvin easterly waves than Rossby westerly waves in the lower troposphere. 

The stationary solution allows the interpretation of  the propagation speed in terms of the intensities of the two adjacent circulation cells. Propagation declines when the MJO reaches a region with east-west symmetry like the maritime continent \cite{suematsu2022changes}. The paper is organized as follows: In the next Section \ref{Sec-CRW} the dynamics of the compressional Rossby waves is described  and in Section \ref{Sec-Burgers} we derive the Burgers equation for a convective event. In Section \ref{Sec-Sum} the results are summarized and discussed.

\section{Compressional Rossby waves} \label{Sec-CRW}

The nontraditional approximation in a vertical plane close to the equator allows a type of linear waves in the horizontal vorticity component $\eta$ with the nontraditional Coriolis term  in the restoring force \cite{ong2020compressional}. Since these waves share the geometric setting with the Burgers equation derived below their dynamic mechanism is recapitulated here. 
The waves are described in the Boussinesq approximation and propagate eastward in the equatorial $x$-$z$-plane  due to the so-called compressible $\beta$-effect, coined in analogy to the $\beta$-effect in the midlatitudes \cite{ong2020compressional}. 

The 3D vortex vector ${\bm \omega}=\nabla \times{\bm u}$ evolves according to the Helmholtz vorticity equation, see for example \cite{Ong-Yang_2022}
\begin{equation} \label{Helmh}
	\frac{d}{dt} {\bm \omega} = {\bm \omega}_a \cdot \nabla {\bm u} - {\bm \omega}_a \nabla \cdot {\bm u}
	+ \nabla \times(b {\bm e}_z), 
	\quad {\bm \omega}_a={\bm \omega}+2{\bm \Omega} , 
%	+{\bf S}, 	\quad 	{\bf S}= \frac{\nabla \rho \times \nabla p}{\rho^2}
\end{equation}
with the absolute vorticity $
{\bm \omega}_a$, the angular rotation rate of the earth ${\bm \Omega}$ and buoyancy $b$.
% , and without the thermodynamic solenoid term ${\bm S}= \nabla \rho \times \nabla p/\rho^2$. 
%

%As in \cite{ong2020compressional} we consider the circulation in a vertical $x$-$z$-plane. 
%
The meridional component of the relative vorticity is $\eta = \omega_y  = \partial_z u - \partial_x w$, 
and the absolute meridional vorticity on the sphere is $\eta_a = \tilde{f} + \eta$ with  the so-called cos-Coriolis parameter $\tilde{f} = 2\Omega \cos \varphi$ at the latitude $\varphi$ (note that $\tilde{f}$ is neglected in TA, see \cite{white1995dynamically}).
Observations of the MJO find  that the meridional velocity component is much smaller than the zonal component, $O(v) \ll O(u)$ \cite{hsu2012role}, hence we use $v=0$ and neglect the tilting term for $\eta$ in (\ref{Helmh}) (first term on the rhs). 

In the Boussinesq approximation density decreases as $\rho \propto \exp(-z/H)$ with a scale height $H=RT/g$. In the moist tropical atmosphere stratification can be considered to be neutral with the Brunt-V\"{a}is\"{a}l\"{a} frequency $N=0$  \cite{ong2020compressional}. 
The flow is anelastic satisfying the vertical continuity equation $\partial_x (\rho u) + \partial_z (\rho w)=0$ for the momentum $(\rho u, \rho w)$, hence a mass stream-function $\Psi$ for the momentum $\rho u = \partial_z \Psi$, $\rho w = - \partial_x \Psi$ is introduced.
The divergence in the vertical plane is $\partial_x u+ \partial_z w=w/H$. 

The first baroclinic mode is written with a  function $\psi$ describing the zonal dependency
\begin{equation} \label{psixzt}
	\Psi(x,z,t) = \psi(x,t) e^{-z/(2H)} \sin(mz), %  \sin(kx-\nu t) , 
\end{equation}
where $m=\pi/H_T$ is the vertical wavenumber with the  tropopause height $H_T$. The vertical exponential decay of $\Psi$ ensures constant vertical kinetic energy, $\rho (u^2+w^2)/2$, with vertically increasing velocities $(u,w) \propto \rho^{-1/2} \propto  e^{z/(2H)}$.
The meridional vorticity $\eta$ is in terms of the mass stream-function 
\begin{equation}
   \rho \eta =  \Psi_{xx}+\Psi_{zz} +\frac{1}{H} \Psi_z ,  
\end{equation}
where subscripts denote derivatives.

% In the following 
% The $z$-dependent prefactor compensates density decrease in energy $\sim \rho (u^2+w^2)$

The meridional vorticity equation at the equator ($\varphi=0$) without buoyancy $b$ is 
% with (\ref{Helmh})
\cite{ong2020compressional}
\begin{equation} \label{etaeqn}
	\frac{\partial \eta}{\partial t} + \frac{\eta_a}{H} w = 0, % + \frac{\partial b}{\partial x} = 0 ,
\end{equation}
with the divergence term in the Boussinesq approximation and without nonlinear advection.
The linearized meridional vorticity equation  is obtained for $\eta  \ll  \Omega$  in $\eta_a$,
\begin{equation} \label{etawlin}
	\frac{\partial \eta}{\partial t} + \hat{\beta} w = 0 ,
\end{equation}
with the compressional $\beta$-parameter $ \hat{\beta} = 2\Omega/H$.

For a wave ansatz with zonal wavenumber $k$ and  frequency $\nu$, 
\begin{equation} \label{Psi}
	\psi_w(x,t) =  \sin(kx-\nu t) , 
\end{equation}
the vorticity is
\begin{equation} \label{etaCCRW}
	\eta = - (k^2+\kappa^2) \Psi, \qquad 
	\kappa^2 =  m^2 + \frac{1}{4H^2} , 
\end{equation}
and the dispersion relation for free waves is obtained,
\begin{equation} \label{nuOR}
	\nu = \hat{\beta} \frac{k}{k^2+\kappa^2} .
\end{equation}
This leads to the zonal phase speed $c=\nu/k$ with $c \approx 0.24 \, \mbox{ms}^{-1}$ for $\Omega=7.292 \cdot 10^{-5} \, \mbox{s}^{-1}, H=9.1 \cdot 1000 \, \mbox{m}$, and $m=2\pi/25000 \mbox{m}^{-1}$ for long zonal waves $k^2 \ll m^2$ \cite{ong2020compressional}. 
Obviously, this mechanism cannot explain the observed MJO speed ($c \approx 5 \, \mbox{ms}^{-1}$).
Since the magnitude of the observed vorticity $O(\eta)  \sim 10^{-3} \, \mbox{s}^{-1}$  is much larger than $O(2\Omega) \sim 10^{-4} \, \mbox{s}^{-1}$ the  equation for finite amplitude compressional Rossby waves becomes nonlinear (here the symbol $\sim$ means equal orders of magnitude or scales).

In the next section we will assume that convection is the major process in the dynamics of the MJO, and concentrate on the corresponding nonlinear mechanism. This leads to a solvable nonlinear equation with a propagating step-like finite amplitude disturbance. 
The compressional Rossby wave mechanism is not pursued further. 

%(note that the nonlinear equation has a similar structure as the Burgers equation below).
 
%This nonlinear equation might be relevant for the propagation of dry disturbances when the vorticity $O(\eta)  \sim 10^{-3} \, \mbox{s}^{-1}$  with $O(2\Omega) \sim 10^{-4} \, \mbox{s}^{-1}$
%alters the dynamics  a scale analysis reveals a meridional vorticity $O(\eta)  \sim 10^{-3} \, \mbox{s}^{-1}$  which becomes the dominant part in the absolute vorticity since $O(2\Omega) \sim 10^{-4} \, \mbox{s}^{-1}$ is much smaller, hence we will not pursue the dynamics of linear compressible Rossby waves.
 
%This means that the compressional vorticity equation (\ref{etawlin}) should contain the term  $   \hat{\beta}_\eta = (2\Omega + \eta)/H $, which renders the equation nonlinear. 

%-------------------------------------------------------------------------------
\begin{figure}[h]
\centering
\includegraphics[angle=0, width=0.45\textwidth]{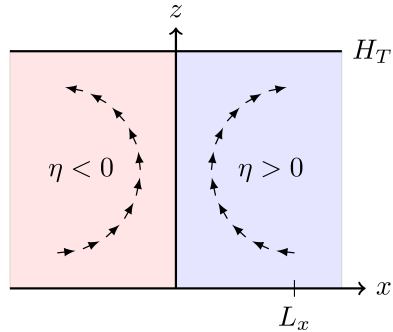}  
\caption{\label{Fig-eta-moist} 
Sketch of the circulation between two adjacent equatorial convection cells in the $x$-$z$-plane with opposite meridional vorticity, $\eta=\partial_z u - \partial_x w$ in terms of the zonal and vertical velocities ($u,w$). $L_x$ is the zonal scale ($10^7$ m) and $H_T$ the equatorial tropopause height (magnitude $2\cdot 10^4$ m). 
}
\end{figure}
%-------------------------------------------------------------------------------

%====================================================
\section{Burgers equation for the MJO progression} \label{Sec-Burgers}
%====================================================

In our model for the  propagation of large scale tropical convection we consider two vorticity cells in the vertical  $x$-$z$-plane with opposite rotation (see Fig. \ref{Fig-eta-moist}). We assume infinitely extended cells and do not specify the remote circulation. 

%Thus the solution captures the local structure.

%Since the eastern and western boundary conditions depend on the remote circulation which is not specified here we assume infinitely extended cells with vorticities approaching constant values. We follow the work on nontraditional compressible Rossby waves propagating with vanishing  meridional velocity $v=0$ \cite{ong2020compressional}. However,  we neglect the compressional $\hat{\beta}$-effect, with $ \hat{\beta} = 2\Omega/H$, since we like to concentrate on the propagation solely based on a convective effect, keeping the model as simple as possible.

To estimate the scale of the meridional vorticity $\eta$ we use  $\eta \sim u/\Delta H \sim 10^{-3} \, \mbox{s}^{-1}$ for the observed zonal velocity scale $u=2.5 \, \mbox{ms}^{-1}$ and a height difference between the 500 and 700 hPa levels, $\Delta H=3000 \, \mbox{m}$ \cite{hsu2012role}.
The linearized meridional vorticity equation  (\ref{Helmh}) with buoyancy $b$ reads as
\begin{equation} \label{etaCCRWb}
	\frac{\partial \eta}{\partial t} =  
	-\frac{\partial b}{\partial x} + D \frac{\partial^2 \eta}{\partial x^2} , 	
\end{equation}
where we neglected the compressible $\beta$-effect  and added a diffusion term. 

In order to parametrize convection in the meridional vorticity equation we use Convective Available Potential Energy (CAPE) which is an indicator for precipitation and thunderstorms \cite{zhang2005effects,narendra2010seasonal}. We  parametrize buoyancy  $b$ in CAPE in terms of the large scale vorticity by a relation to vertical kinetic energy. A moisture source is located in the east of the MJO (in the maritime continent) and moisture is transported by near surface advection with westward zonal velocity $u<0$.

For vertically constant buoyancy CAPE is $h_c b(x,t) =\hat{w}^2/2$ with maximum vertical velocity $\hat{w}(x,t)$, and the height difference $h_c  \sim 5 \cdot10^{3} \, \mbox{m}$ between the levels of neutral buoyancy and free convection. With typical CAPE values up to $\sim 500 \, \mbox{Jkg}^{-1}$ in the tropics we obtain $ b \sim 0.1 \, \mbox{ms}^{-2}$ and a moderate estimate  $\hat{w} \sim 10 \, \mbox{ms}^{-1}$. The horizontal derivative $b_x$ in the vorticity equation (\ref{etaCCRWb}) could be written as $b_x \sim (1/h_c) \hat{w} \hat{w}_x$, however, large scale observed vertical velocities are of the order of $w \sim 10^{-3} \, \mbox{ms}^{-1}$ \cite{hsu2012role}, much lower than the maximum values in CAPE.    

The parametrization is closed by a relation between the large scale vertical velocity and $\eta$.  The scales assumed here are the zonal extent $L_x = 10^{7} \, \mbox{m}$ of the MJO event with a wavenumber $k = 6.28 \cdot 10^{-7} \, \mbox{m}^{-1}$, the time scale $T=10^6 \, \mbox{s}$, and the meridional vorticity $\eta \sim 10^{-3} \, \mbox{s}^{-1}$. For the parametrization we assume a wave-like structure in the vicinity of the MJO center such that $\eta_{xx} \approx -k^2 \eta$. Thus the solution for the vorticity determined below is an envelope which decays on a scale only weakly larger than $k^{-1}$. For  $\kappa^2  = m^2 = 6.38 \cdot 10^{-8} \, \mbox{m}^{-2}$ we use the values in (\ref{nuOR}).
%$\partial_t \sim 1/T$, $\partial_x \sim 1/L_x$,  %$k$: $6.6 \cdot 10^{-8} \, \mbox{m}^{-1}$, 

The vertical velocity is $w = \kappa^{-2} \eta_x$, and  $w_x = \kappa^{-2} \eta_{xx} =  -k^2 \kappa^{-2} \eta$ (\ref{etaCCRW}).  This yields $w = 1.58 \cdot 10^{-3} \, \mbox{ms}^{-1}$, which is in the range of the large scale observations \cite{hsu2012role}. These relations suggest a parametrization for the zonal buoyancy derivative in terms of the meridional vorticity with the functional form $b_x \sim -(k^2 \kappa^{-4}/h_c) \eta \eta_x$. 

However, due to the huge gap between different scales and different processes involved, the observed large scale vertical velocity is much lower than the observed maximum value $\hat{w}$ in CAPE. Therefore, we resort to an empirical condition with a length scale $s$,
\begin{equation} \label{seta}
    s \eta^2 = \frac{1}{2 h_c} \hat{w}^2.
    %\sim -(1/h_c) (k^2/ \kappa^{4}) \eta \eta_x .   
 \end{equation}
With the scales above this yields $s = 10^{4} \, \mbox{m}$, comparable to the half of the tropopause height, $s \sim H_T/2$.   Finally, we parametrize buoyancy in (\ref{etaCCRWb}) as 
\begin{equation} \label{bx}
    \frac{\partial b}{\partial x} = -s \eta \frac{\partial \eta}{\partial x} .    
 \end{equation}
The dynamics is  effectively one-dimensional for the relative vorticity $\eta(x,t)$ with the vertical structure given by the first baroclinic mode as in (\ref{psixzt}),
\begin{equation}
	\hat{\eta}(x,z,t) = \eta(x,t) e^{z/(2H)} \sin(mz) .
\end{equation}
Equation  (\ref{bx}) allows a heuristic interpretation of the parametrized convection in terms of moisture transport with the azimuthal velocity in a vertical distance $s$ around a vorticity center $\eta$, $u \sim  s \eta/2$, with the angular rotation rate $\eta/2$ in the vertical plane.    Additionally, the vertical velocity $\eta_x$ has to be positive in convection.  

Thus, an advantage of the nonlinear parametrization is that the sign of the moisture transport is mimicked by the sign of $s\eta$. For the signs of $\eta$ in Fig.~\ref{Fig-eta-moist} the moisture source region can be associated with the sign of $s$, as for an easterly (westerly) moisture source we have a scale $s>0$ ($s<0$) which leads to positive advection of moisture in the eastern (western) cell by $\eta>0$ ($\eta<0$). 
The  solution will show that the cell propagates towards the moisture source (eastward for $s>0$, and westward for $s<0$). 

The propagation of an initial spontaneous convection cell with convection in the center can be understood as follows: For a wave-like form $\eta \sim \sin(kx)$, the buoyancy term $\eta \eta_x$ is positive in a region east of the center. This causes a phase shift between buoyancy and circulation which drives the cell eastward. 

To maintain a stationary solution a dissipative process is required which is described here as a linear diffusion term $D\eta_{xx}$. We apply the diffusion coefficient $D = 10^{8} \, \mbox{m}^2\mbox{s}^{-1}$  which corresponds to a mean-square displacement scaling with $L_x^2/T$, corresponding to the magnitude of the vorticity tendency and the buoyancy forcing. A plausible process is the energy loss of the large scale circulation by the radiation of gravity waves as observed in the vicinity of tropical storms \cite{Jewtoukoff_Gravitywaves_2013} (note that the diffusion coefficients therein are much smaller). 

% with wave lengths  $\ell \sim 10^5 \, \mbox{m}$, periods $\tau \sim 10^3 \, \mbox{s}$,  and average phase speeds  $c_{ph} \sim \ell/\tau \sim 25 \, \mbox{ms}^{-1}$. Note that this speed agrees with $L_x/T$. 

With the parametrization (\ref{bx}) the meridional vorticity equation (\ref{etaCCRWb}) becomes the Burgers equation,
\begin{equation} \label{Burgerseq}
	\frac{\partial \eta}{\partial t} 
	-  s \eta \frac{\partial \eta}{\partial x} 
	= D \frac{\partial^2 \eta }{\partial x^2} .
\end{equation}
Solutions of the Burgers equation are stationary if nonlinear steepening is balanced by diffusion (in contrast to the Korteweg-de Vries equation where steepening is balanced by dispersion, see e.g. \cite{khouider2012climate}). 

The Burgers equation (\ref{Burgerseq}) allows a solution (see e.g. \cite{benton1972table, bonkile2018systematicBurgers}),
\begin{equation} \label{kink}
	\eta = -\frac{c}{s} +\frac{2 \lambda D}{s} \tanh(\lambda(x-ct) ) , 
\end{equation}
which describes a kink with two free parameters, the propagation speed $c$ and an arbitrary inverse width $\lambda$ which also determines the amplitude. 
This exact nonlinear solution is the main advantage of the present model since it reveals an unsuspected relation between  the propagation speed and the intensities of the circulation cells in the MJO. 
The solution incorporates the essential constant mean vorticity  $\bar{\eta} = -c/s$ which is considered in more detail below. 

%Below we will assume $\lambda=1/L_x = 10^{-7} \, \mbox{m}^{-1}$. 
With $s=-1$, (\ref{Burgerseq}) can be brought in standard form $U_t + UU_x = D U_{xx}$. 
In this case $U$ represents the zonal velocity in the lower troposphere.
Exact solutions can be obtained by the Hopf Cole transformation $U = -2D \partial_x \log \Phi$ where  $\Phi$ satisfies a linear diffusion equation.

%Further solutions can be found for example in \cite{benton1972table}. 
% (exact solutions like (\ref{kink}) can be found easily with symbolic iteration \cite{blender1991iterative}). 

\begin{figure}[h]
\centering
\includegraphics[angle=-90, width=0.7\textwidth]{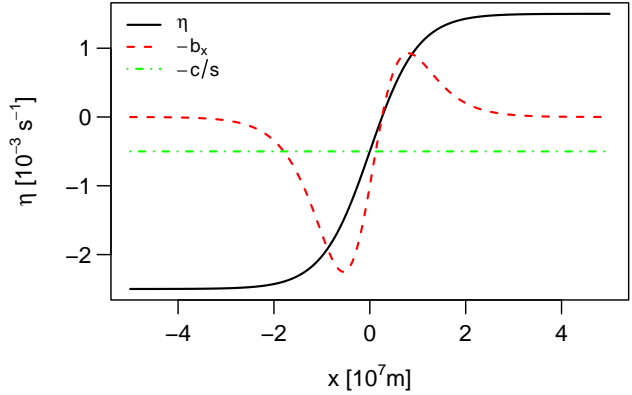}  
\caption{\label{Fig-MJO-kinkheat1} 
Burgers kink solution for $\eta$ (Eq. \ref{kink}, solid) with the nonlinear buoyancy term $-b_x$ (Eq. \ref{bx}, dashed), and the constant term $-c/s$ (dash-dotted), which causes the east-west asymmetry. The parameters are: observed propagation speed $c=5 \, \mbox{ms}^{-1}$,  scale  $s = 10^4 \, \mbox{m}$, diffusion coefficient $D = 10^8 \, \mbox{m}^2 \mbox{s}^{-1}$,  
and inverse width $\lambda=1/L_x=10^{-7} \, \mbox{m}^{-1}$. 
The displayed functions are nondimensional with the scale $10^{-3} \, \mbox{s}^{-1}$.
}
\end{figure}

The solution (\ref{kink}) describes a kink, a smooth step of the meridional vorticity $\eta$,  see  Fig. \ref{Fig-MJO-kinkheat1}. This kink describes the circulation in the vicinity of an MJO event with a moisture source in the east. 
In the eastern and western directions the vorticity converges to constant values. 
According to the parametrization of buoyancy in terms of CAPE, precipitation is predicted in the east for $s>0$ close to the origin where $b_x$ is negative (\ref{bx}). 

The buoyancy  term $-b_x$ causes nonlinear steepening since it enhances positive vorticity in the east and negative vorticity in the west (Fig. \ref{Fig-MJO-kinkheat1}). In this process the vertical velocity ($ \propto \eta_x$) in the center increases and the convective region narrows. The curve is stationary since steepening is compensated by horizontal diffusion. An MJO envelope can be identified by $\eta_x$.

The Burgers equation (\ref{Burgerseq}) can be rewritten with a total derivative following the MJO event if the vorticity is decomposed in a constant term $\bar{\eta}=-c/s$ and a localized anomaly  $\eta^\prime$, $\eta=\bar{\eta}+ 
\eta^\prime$,
%
%\begin{equation} \label{etaq}
%	\bar{\eta} = -\frac{c}{s}, 
%	\quad \eta^\prime = \frac{2 \lambda D}{s} \tanh(\lambda(x-ct) ).
%\end{equation}
%
% Thus (\ref{Burgerseq}) can be written with advection and nonlinear interaction terms 
% 
\begin{equation} \label{etacq}
	\frac{D}{Dt} \eta^\prime
	\equiv \left( \frac{\partial}{\partial t} +c \frac{\partial}{\partial x}\right) \eta^\prime  
	= s \eta^\prime \frac{\partial}{\partial x} \eta^\prime
	+ D \frac{\partial^2}{\partial x^2}	\eta^\prime.
\end{equation}
The split of $\eta$ in $\bar{\eta}$ and $\eta^\prime$ decomposes the inviscid dynamics in progression (with $\bar{\eta}$) and nonlinear steepening (with $\eta^\prime$). This explains the asymmetric structure of the buoyancy term $-b_x$ in Fig. \ref{Fig-MJO-kinkheat1} with the  term $s \bar{\eta} \eta_x^\prime=-c \eta_x^\prime < 0$, which determines propagation. 
%Equation (\ref{etacq}) shows that the speed is simply given by the mean vorticity  $\bar{\eta} = -c/s$. 

For the comparison with observations the asymmetry of the eastern and western limits 
 $\eta_E = \lim_{x\to \infty} \eta, \eta_W= \lim_{x\to -\infty} \eta$
in (\ref{kink}) is relevant since it is directly linked to the propagation speed,   
\begin{equation}  \label{cetas}
   c=-\frac{s}{2} (\eta_E+\eta_W). 
\end{equation}
%
%where $\eta_E = \lim_{x\to \infty} \eta, \eta_W= \lim_{x\to -\infty} \eta$.
Eastward propagation in the Indian ocean requires $s>0$. 
%This relation indicates the relevance of the sign of the scale $s$ in the propagation direction. 
We assume that both limits correspond to the observed maxima in the adjacent circulation cells.

The lack of a point symmetry of the solution (\ref{kink}) at the origin with a lower absolute $\eta$ value in the east compared to  the west is known from observations \cite{hsu2012role} where the 700 hPa zonal winds  in the east are weaker than in the west (their Figure 13g), $u_W \approx 3 \, \mbox{ms}^{-1}, u_E \approx -2 \, \mbox{ms}^{-1}$. 
Our model determines the winds at the lower boundary as $u_s  \approx -m \kappa^{-2} \eta$ with $m \kappa^{-2} = 3.97 \cdot 10^3 \, \mbox{m}$ and overestimates the observed surface velocities measured below the 700 hPa level.   The counterintuitive result described in \cite{suematsu2022changes} as a slower MJO for an intensified Walker circulation might be explained by higher $\eta_E$ leading to weaker asymmetry and a lower speed $c$. 

A zonal asymmetry is already predicted in \cite{gill1980some} by linear wave theory for a stationary heat source. Note that since this heat source does not propagate a relation between speed and wave intensity is not available. The present model suggests that the heat source propagates due to the zonal asymmetry and drags the linear waves. 

In \cite{Wang-MJOPropagation} the Rossby-Kelvin intensity index (denoted RK here) is defined as the relative strength of the Rossby versus Kelvin wave surface wind velocities and written as the ratio of the maximum westerly speed $U_{max}$ versus the minimum easterly speed $U_{min}$. The zonal asymmetry is found to be a major ingredient in realistic GCM simulations of the MJO propagation.  

We suggest that the RK index is related to the propagation speed $c$ due to the zonal surface wind $u_s$ in the meridional circulation cells. Using $\eta \propto -u_s$ in (\ref{cetas}) the RK index is given by the ratio of eastern to western meridional vorticities $\eta_E$ and $\eta_W$,
\begin{equation}  \label{RKcetas}
   \mbox{RK} = \eta_E/\eta_W .
\end{equation}
The RK index is negative since the vorticities have opposite signs.

For the values indicated in Fig. \ref{Fig-MJO-kinkheat1}, $\eta_W= -2.5 \cdot 10^{-3} \, \mbox{s}^{-1}$, $\eta_E= 1.5 \cdot 10^{-3} \, \mbox{s}^{-1}$, hence  $\mbox{RK}=-0.6$, which is above the observed value $-0.8$ but in the range between $-0.6$ and $-2.4$ simulated by GCMs  \cite{Wang-MJOPropagation}. The speed in (\ref{cetas}) is linearly related to the RK index according to
\begin{equation} \label{cRK}
   c= -\frac{s  \eta_w}{2}(1+\mbox{RK}) .
\end{equation} 
The speed is positive here since $\eta_w<0$ and $\mbox{RK}>-1$. An unrealistic westward propagation of the MJO is given for $\mbox{RK}<-1$. 

%The observed speed $c = 5 \, \mbox{m}\mbox{s}^{-1}$ is obtained for $s = 10^4 \, \mbox{m}$, $\eta_W= -2.5 \cdot 10^{-3} \mbox{s}^{-1}$, and  for the index $\mbox{RK}=-0.6$, which is above the observed value $-0.8$.

%To quantify the asymmetry with a parameter which is independent of the absolute values we define 
%\begin{equation} \label{asym}
%   a= \frac{\eta_E+\eta_w}{\eta_E- \eta_w} = \frac{1+\mbox{RK}}{1-\mbox{RK}}.
%\end{equation}
%\noindent 
%The eastern and western vorticities $\eta_E$ and $\eta_w$ can be replaced by corresponding zonal surface velocities $u_E$ and $u_W$ provided the relation between $u$ and $\eta$ is linear. 
% the large scale horizontal surface velocity is  $u \sim 3.97 \, \mbox{ms}^{-1}$.
%For the kink solution (\ref{kink}) the asymmetry parameter is $a= -cL_x/(2D)=-1/4$ which is in the range of observations  \cite{hsu2012role}. 

%-------------------------------------------------------------------------------
%-------------------------------------------------------------------------------

%Finally a remark on the boundary conditions should be added. 
%The solution describes a kink which can be associated with the MJO event and varies on a length scale $L_x \sim 10^7 \, \mbox{m}$.  However the solution is embedded in large cells, mainly given by the climatological Walker circulation and variations caused bei ENSO on the interannual time scale.  Conditions outside It becomes constant outside satisfies boundary conditions $\eta (x\rightarrow \pm \infty \rightarrow \mbox{const}$ are not specified 

%====================================================
\section{Summary and Discussion} \label{Sec-Sum}
%====================================================

A conceptual mechanism for the propagation of a large scale convective event like the Madden-Julian Oscillation (MJO) in the Indian Ocean is suggested. The dynamics is determined by the meridional vorticity equation in the vertical $x$-$z$-plane in the Boussinesq approximation. Buoyant convection is parametrized resorting to Convective Available Potential Energy (CAPE) in terms of the vorticity and its horizontal derivative, thus it can be considered heuristically as the product of the moisture transport times vertical velocity. The product leads to a phase shift in the buoyancy which causes the eastward progression. Clearly, the model depends crucially on this purely mechanistic parametrization of convection which allows to describe the phenomenon by a single one-dimensional equation, whose usefulness is demonstrated a posteriori.

The evolution equation for the first baroclinic mode is the one-dimensional nonlinear Burgers equation where nonlinear steepening is balanced by horizontal diffusion. The exact stationary solution is a self-sustained perturbation in the form of a smooth step (kink) in the vorticity propagating towards the moisture source with a speed proportional to the mean meridional vorticity in the vicinity of the MJO, $\bar{\eta} \propto -c$.  Due to its large scale structure the kink solution should be considered as an envelope for the meridional vorticity. To establish a connection with equatorial wave theory, the kink could be considered as a slowly moving (quasi stationary) heat source in Gill's shallow water model \cite{gill1980some}.

A first important property of the Burgers kink solution is the zonal asymmetry of the meridional vorticity with lower absolute eastern than western values,  $\eta_E < -\eta_W$, hence a point symmetry is absent in the exact solution. This asymmetry is known in the observed lower tropospheric zonal wind \cite{hsu2012role}. We hypothesize that slower propagation complies with a more zonally symmetric lower boundary, for example when the MJO reaches the maritime continent \cite{hudson2023role}. 

In \cite{Wang-MJOPropagation} the zonal asymmetry is attributed to stronger eastward surface winds in Kelvin waves than westward winds in Rossby waves and assessed by a Rossby-Kelvin intensity index which is  $\mbox{RK} = \eta_E/\eta_W$ in terms of eastern and western meridional vorticities.
% the RK index is given by, and the kink solution shows that the asymmetry is directly related to the propagation speed due to the term $-c/s$.
Since  \cite{gill1980some} obtained the zonal asymmetry in the surface wind for a stationary heat source, the observed relation between the RK index and the MJO speed might be additionally affected by the circulation cells considered in the present study.

A second important result of the present conceptual model is the predicted convective precipitation in a narrow equatorial band eastward of the MJO (for an easterly moisture source) which is not correlated with Kelvin and Rossby waves \cite{zhang2012potential}. 
%Clearly, the nearby reason is that both wave types rely on the quasigeostrophic approximation which becomes invalid at the equator. 
The eastward location of precipitation can be associated with positive values of the buoyancy term $-b_x$ 
%if we assume a moisture source in the east 
(compare also the asymmetric structure of convective instability in Fig. 4 of \cite{Wang-MJOPropagation}).   

%It is possible that the combination of the present model with a Gill type model for a progressing heat source as in \cite{kacimi2018transient} revels a more complete circulation. 

The present model puts the MJO in the framework of circulation cells like the Walker circulation and  
 the El Ni\~no/Southern Oscillation (ENSO) \cite{zhang2005madden} which can be described in terms of a meridional vorticity as well.  Therefore, interactions are straightforward, for example, the meridional vorticity and the MJO are weakened when the upward motion over the Maritime Continent decreases during a warm ENSO event, the opposite happens for an intensified Walker circulation \cite{suematsu2022changes,suematsu2022deceleration}. In \cite{nishimoto2013intraseasonal} the propagation speeds of convective clusters in the Indian Ocean/West Pacific are classified for different ENSO phases. 

Presumably, the most important result of the present study is not the Burgers equation itself, but the exact kink solution which associates the speed of the kink with an asymmetry of the zonal circulation in the vicinity of the kink. Remote boundary conditions for the solution are not specified here but might be determined by planetary circulation cells like the Walker circulation and the anomaly given by ENSO. These cells are defined or can be interrupted by large scale regions of ascending and descending motion which appear as smooth discontinuities (kinks) in the meridional vorticity. The model predicts that kinks associated with convection move towards the moisture source and break the symmetry of the circulation cells locally, while kinks in descending regions or without convection are symmetric and do not move.

%On the other hand, ENSO interferes with teleconnections of the MJO (e.g. rainfall in south Africa, see \cite{pohl2007influence}).

%====================================
\section*{References}

\bibliographystyle{unsrt}

\bibliography{Blender-MJO-references}	 

\end{document}